\documentclass[preprintnumbers,superscriptaddress,showpacs,pra]{revtex4}
\usepackage{amssymb,amsmath}
\usepackage{graphicx}
\usepackage{amsmath}
\usepackage{amsfonts}

\input{tcilatex}
\begin{document}

\preprint{}
\title[Cartesian momentum operators]{Constraint induced mean curvature
dependence of Cartesian momentum operators}
\author{Q. H. Liu}
\affiliation{School for Theoretical Physics, and Department of Applied Physics, Hunan
University, Changsha, 410082, China}
\author{C. L. Tong}
\affiliation{School for Theoretical Physics, and Department of Applied Physics, Hunan
University, Changsha, 410082, China}
\author{M. M. Lai}
\affiliation{School for Theoretical Physics, and Department of Applied Physics, Hunan
University, Changsha, 410082, China}
\date{\today}

\begin{abstract}
The Hermitian Cartesian quantum momentum operator $\mathbf{p}$ for an
embedded surface $M$\ in $R^{3}$ is proved to be a constant factor $-i\hbar $
times the mean curvature vector field $H\mathbf{n}$ added to the usual
differential term. With use of this form of momentum operators, the
operator-ordering ambiguity exists in the construction of the correct
kinetic energy operator and three different operator-orderings lead to the
same result.
\end{abstract}

\maketitle

\section{Introduction}

For a particle moves on the curved smooth (regular) surface $M$ embedded\ in 
$R^{3}$, which is parameterized by two local coordinates ($\xi ,\zeta $),
the quantum kinetic energy operator takes the following form,%
\begin{equation}
T\equiv -\frac{\hbar ^{2}}{2m}\nabla ^{2}=-\frac{\hbar ^{2}}{2m}\frac{1}{%
\sqrt{g}}\partial _{\mu }g^{\mu \upsilon }\sqrt{g}\partial _{\upsilon }\text{%
,}  \label{standard}
\end{equation}%
where 
\begin{equation}
\nabla ^{2}=\partial _{i}\partial _{i}=\frac{1}{\sqrt{g}}\partial _{\mu
}g^{\mu \upsilon }\sqrt{g}\partial _{\upsilon }  \label{lap-bel}
\end{equation}%
is the Laplace-Beltrami operator \cite{kleinert}. The symbol $\partial $
stands for differential operator as usual. The metric tensor $g_{\mu
\upsilon }$ is defined via the length element square $ds^{2}=$\ $g_{\mu
\upsilon }dx^{\mu }dx^{\upsilon }$ and $d\sigma =\sqrt{g}d\xi d\zeta $ is
the area element on the surface. The factor $g\equiv \det (g_{\mu \upsilon
}) $ is the determinant of the matrix formed by the metric tensor. In this
paper the Latin indices $(i,j,k)$ are used to denote the Cartesian
coordinates $\mathbf{(}x,y,z)$ with $x^{i}=x_{i}$ and Greek indices $(\mu
,\upsilon )$ to denote the local ones $(\xi ,\zeta )$ with $x^{\mu }=g^{\mu
\upsilon }x_{\nu }$. The convention the repeated indices mean summation is
implied unless specified. Only two-dimensional surface embedded in the
three-dimensional Euclidean space is addressed in this paper because in
majority of the realistic constraint problems, the motion is on the
two-dimensional curved surfaces \cite{nanophys1, nanophys2}. However, our
conclusion can be readily generalized to the higher-dimensional manifold.

For the constraint motion, the quantum kinetic energy operator can be
rewritten into a form depending on the \textit{generalized momentum operators%
} $p_{\mu }$ as \cite{kleinert},%
\begin{equation}
T\equiv \frac{1}{2m}\frac{1}{g^{1/4}}p_{\mu }g^{1/4}g^{\mu \upsilon
}g^{1/4}p_{\upsilon }\frac{1}{g^{1/4}}  \label{curved}
\end{equation}%
where the generalized momentum operators $p_{\mu }$ $(\mu =\xi ,\zeta )$ are
with $\Gamma _{\mu }\equiv \Gamma _{\mu \upsilon }^{\upsilon }$ being the
once-contracted affine connection, 
\begin{equation}
p_{\mu }=-i\hbar (\partial _{\mu }+1/2\Gamma _{\mu }).  \label{canP}
\end{equation}%
In the kinetic energy (\ref{curved}), the four identical $g^{1/4}$ factors
are used to fix the operator-ordering problem, and they are so inserted that
the standard result (\ref{standard}) can be restored. In classical limit,
these factors drop out and Eq. (\ref{curved}) becomes, 
\begin{equation}
T\equiv \frac{1}{2m}g^{\mu \upsilon }p_{\mu }p_{\upsilon }.  \label{classT}
\end{equation}

Similarly, when examining the same constraint motion in Cartesian
coordinates with use of the Hermitian form of \textit{Cartesian momentum} $%
p_{i}$ $(i=x,y,z)$, the elaboration of the kinetic energy operator should
also take appropriate account of the operator-ordering problem. In analogy
of (\ref{curved}) the quantum kinetic energy operator may take the following
form, 
\begin{equation}
T=\frac{1}{2m}\overset{3}{\underset{i=1}{\sum }}\sum\limits_{i=1}^{3}\frac{1%
}{f_{i}}p_{i}f_{i}^{2}p_{i}\frac{1}{f_{i}},  \label{general}
\end{equation}%
where the Cartesian momentum $p_{i}$ depend on two independent curved
coordinates $(\xi ,\zeta )$ and their first derivatives only, and the
operator-ordering factors $f_{i}$ $(i=x,y,z)$ are non-trivial functions
depending on the local coordinates $(\xi ,\zeta )$ too. When the constraint
is removed or the motion is in classical limit, the factors $f_{i}(x,y,z)$
cancel out; and the kinetic energy operator (\ref{general}) reduces to be
its usual form,%
\begin{equation}
T=\frac{1}{2m}p_{i}p_{i}.  \label{flatT}
\end{equation}%
It can be anticipated that the Hermitian form of the Cartesian quantum
momentum operators $p_{i}$ may take a form similar to (\ref{canP}), which
proves to be,%
\begin{equation}
p_{i}=-i\hbar (\partial _{i}+Hn_{i}),  \label{flatP}
\end{equation}%
where $H$ is the \textit{mean curvature} of the surface $M$ in which $%
\mathbf{n}=(n_{x},n_{y},n_{z})$ denoting the unit normal vector on the
surface, and the quantity $H\mathbf{n}$ is an existing \textit{geometric
invariant} in differential geometry, the so-called \textit{mean curvature
vector field} \cite{smirnov}.

This paper is organized as what follows. A proof of result (\ref{flatP}) is
given in Section II. The condition for the operator-ordering factors $f_{i}$
being able to convert Eq. (\ref{general}) into Eq. (\ref{standard}) is
derived in Section III, which is found to depend on the mean curvature $H$
also. However, the way of inserting $f_{i}$ into $p_{i}p_{i}$ (\ref{flatT})
is not unique, and two other ways of the insertion can yield the correct
result (\ref{standard}), as shown in Section IV. To illustrate the abstract
formulae obtained, the explicit results for two surfaces are given in
Section V. In final Section VI, some remarks are provided.

\section{Hermitian Cartesian quantum momentum operator}

The standard representation of the curved smooth surface $M$ embedded\ in $%
R^{3}$ is,%
\begin{equation}
\mathbf{r}(\xi ,\zeta )\mathbf{=}\left( x(\xi ,\zeta ),y(\xi ,\zeta ),z(\xi
,\zeta )\right) .  \label{R}
\end{equation}%
The covariant derivatives\ of $\mathbf{r}$ (\ref{R})\ are $\mathbf{r}_{\mu
}=\partial \mathbf{r}/\partial x^{\mu }$, and then the metric tensor $g_{\mu
\upsilon }$ is easily formed as $g_{\mu \upsilon }\equiv \mathbf{r}_{\mu
}\cdot \mathbf{r}_{\upsilon }$. The normal vector at point $(\xi ,\zeta )$
is $\mathbf{n=r}^{\xi }\times \mathbf{r}^{\zeta }/\sqrt{g}$. The
contravariant derivatives $\mathbf{r}^{\mu }\equiv g^{\mu \upsilon }\mathbf{r%
}_{\upsilon }$ is the generalized inverse (or pseudoinverse, or
Moore-Penrose inverse) of the covariant ones $\mathbf{r}_{\mu }$ for we have 
$\mathbf{r}^{\mu }\cdot \mathbf{r}_{\upsilon }=g^{\mu \alpha }\mathbf{r}%
_{\alpha }\cdot \mathbf{r}_{\upsilon }=g^{\mu \alpha }g_{\alpha \upsilon
}=\delta _{\upsilon }^{\mu }$. The derivatives $\mathbf{r}^{\mu }$ and $%
\mathbf{r}_{\upsilon }$ actually constitute the transformation matrix
between $\partial _{i}$ and $\partial _{\mu }$, and explicitly we have, 
\begin{equation}
\partial _{i}=x_{i}^{\mu }\partial _{\mu }\text{, and }\partial _{\mu
}=x_{\mu }^{i}\partial _{i}.  \label{coortran}
\end{equation}%
In consequence, operator $\partial _{i}\partial _{i}=x_{i}^{\mu }\partial
_{\mu }x_{i}^{\upsilon }\partial _{\upsilon }$ is the Laplace-Beltrami
operator (\ref{lap-bel}) for the surface,%
\begin{equation}
\partial _{i}\partial _{i}=x_{i}^{\mu }\partial _{\mu }x_{i}^{\upsilon
}\partial _{\upsilon }=\mathbf{r}^{\mu }\partial _{\mu }\cdot \mathbf{r}%
^{\upsilon }\partial _{\upsilon }=g^{\mu \upsilon }\partial _{\mu }\partial
_{\upsilon }-\Gamma _{\mu }^{\mu \upsilon }\partial _{\upsilon }=\nabla ^{2},
\label{laplacian}
\end{equation}%
where the Gauss formula $\partial _{\mu }\mathbf{r}^{\upsilon }=-\Gamma
_{\gamma \mu }^{\upsilon }\mathbf{r}^{\gamma }+b_{\mu }^{\upsilon }\mathbf{n}
$ \cite{smirnov} is used. Using the Bohm's rule \cite{bohm}, we obtain the
Hermitian form of the operators $-i\hbar \partial _{i}$, and it is,%
\begin{eqnarray}
p_{i} &\equiv &\frac{1}{2}\left\{ (-i\hbar \partial _{i}+(-i\hbar \partial
_{i})^{\dag }\right\}  \notag \\
&=&-i\hbar \left\{ x_{i}^{\mu }\partial _{\mu }+\frac{1}{2\sqrt{g}}\partial
_{\mu }(\sqrt{g}x_{i}^{\mu })\right\}  \notag \\
&=&-i\hbar \left\{ x_{i}^{\mu }\partial _{\mu }+H_{i}\right\} ,(i=1,2,3),
\label{hermitian}
\end{eqnarray}%
where 
\begin{equation}
H_{i}\equiv \frac{1}{2\sqrt{g}}\partial _{\mu }(\sqrt{g}x_{i}^{\mu })
\label{Hi}
\end{equation}%
is the constraint induced term. Rewriting (\ref{Hi}) into the vector form,
we see, 
\begin{equation}
\mathbf{H}\equiv \frac{1}{2\sqrt{g}}\partial _{\mu }(\sqrt{g}\mathbf{r}^{\mu
})=\frac{1}{2\sqrt{g}}\partial _{\mu }(\sqrt{g}g^{\mu \upsilon }\partial
_{\nu }\mathbf{r})=\frac{1}{2}\nabla ^{2}\mathbf{r}=H\mathbf{n.}
\end{equation}%
In last step, the formula $\nabla ^{2}\mathbf{r}=2H\mathbf{n}$ \cite{oy} is
used. For those who are unfamiliar with this formula, another
straightforward proof is available. Recalling the Gauss formula $\partial
_{\upsilon }\mathbf{r}^{\mu }=-\Gamma _{\gamma \upsilon }^{\mu }\mathbf{r}%
^{\gamma }+b_{\upsilon }^{\mu }\mathbf{n}$ \cite{smirnov} and using two
relations $\Gamma _{\mu \upsilon }^{\upsilon }=\partial _{\mu }\ln \sqrt{g}$
and $b_{\mu }^{\mu }\equiv g^{\mu \upsilon }b_{\mu \upsilon }=2H$ \cite%
{smirnov}, we have for $\mathbf{H}$, 
\begin{equation}
\mathbf{H}=\frac{1}{2}(\partial _{\mu }\mathbf{r}^{\mu }+\Gamma _{\mu
\upsilon }^{\upsilon }\mathbf{r}^{\mu })=\frac{1}{2}(-\Gamma _{\mu \upsilon
}^{\upsilon }\mathbf{r}^{\mu }+b_{\mu }^{\mu }\mathbf{n+}\Gamma _{\mu
\upsilon }^{\upsilon }\mathbf{r}^{\mu })=H\mathbf{n.}
\end{equation}%
Thus, the Hermitian Cartesian momentum\ $\mathbf{p}$ (\ref{hermitian}) is in
its final form,%
\begin{equation}
\mathbf{p=}-i\hbar (\mathbf{r}^{\mu }\partial _{\mu }+H\mathbf{n).}
\label{p}
\end{equation}%
When the motion is constraint-free or in a flat plane, i.e., when $H=0$, the
constraint induced terms $H\mathbf{n}$ vanish. Then the Cartesian momentum
operator (\ref{p})\ reproduces its usual form as,%
\begin{equation}
\mathbf{p=}-i\hbar \nabla .
\end{equation}

\section{Kinetic operator in terms of the Hermitian Cartesian momentum
operators}

With use of the Hermitian form of momentum operator (\ref{p}), the correct
kinetic energy operator can no longer be expressed by, 
\begin{equation}
T=\frac{1}{2m}(p_{x}^{2}+p_{y}^{2}+p_{z}^{2}),
\end{equation}%
which will be shortly seen to include an excess positive term $(\hbar
^{2}/2m)H^{2}$ in comparison with the correct kinetic operator (\ref%
{standard}),%
\begin{equation}
\frac{1}{2m}(p_{x}^{2}+p_{y}^{2}+p_{z}^{2})=-\frac{\hbar ^{2}}{2m}\nabla
^{2}+\frac{\hbar ^{2}}{2m}H^{2}.  \label{wrong}
\end{equation}%
So, the operator-ordering problem must be taken into consideration, and we
can resort to the form of Eq. (\ref{general}). Substituting $p_{i}$ (\ref{p}%
) into Eq. (\ref{general}), we have,%
\begin{eqnarray}
T &=&\frac{1}{2m}\overset{3}{\underset{i=1}{\sum }}\frac{1}{f_{i}(x,y,z)}%
p_{i}f_{i}(x,y,z)f_{i}(x,y,z)p_{i}\frac{1}{f_{i}(x,y,z)}  \notag \\
&=&-\frac{\hbar ^{2}}{2m}\overset{3}{\underset{i=1}{\sum }}(\frac{1}{f_{i}}%
x_{i}^{\mu }\partial _{\mu }f_{i}+H_{i})(f_{i}x_{i}^{\upsilon }\partial
_{\upsilon }\frac{1}{f_{i}}+H_{i})  \notag \\
&=&-\frac{\hbar ^{2}}{2m}\overset{3}{\underset{i=1}{\sum }}(x_{i}^{\mu
}\partial _{\mu }+H_{i}+x_{i}^{\mu }(\partial _{\mu }\ln
f_{i}))(x_{i}^{\upsilon }\partial _{\upsilon }+H_{i}-x_{i}^{\upsilon
}(\partial _{\upsilon }\ln f_{i}))  \notag \\
&=&-\frac{\hbar ^{2}}{2m}(x_{i}^{\mu }\partial _{\mu
}+H_{i}+R_{i})(x_{i}^{\upsilon }\partial _{\upsilon }+H_{i}-R_{i}),
\label{T}
\end{eqnarray}%
where 
\begin{equation}
R_{i}\equiv x_{i}^{\mu }(\partial _{\mu }\ln f_{i})\text{, \ (no summation
over two repeated indices }i\text{).}  \label{hdepend}
\end{equation}%
Expanding the right hand side of Eq. (\ref{T}), we find,%
\begin{eqnarray}
T &=&-\frac{\hbar ^{2}}{2m}(x_{i}^{\mu }\partial _{\mu }x_{i}^{\upsilon
}\partial _{\upsilon }+x_{i}^{\mu }\partial _{\mu }H_{i}-x_{i}^{\mu
}\partial _{\mu }R_{i}+H_{i}x_{i}^{\upsilon }\partial _{\upsilon
}+R_{i}x_{i}^{\upsilon }\partial _{\upsilon }+H_{i}^{2}-R_{i}^{2})  \notag \\
&=&-\frac{\hbar ^{2}}{2m}(x_{i}^{\mu }\partial _{\mu }x_{i}^{\upsilon
}\partial _{\upsilon }+\left\{ 2H_{i}x_{i}^{\mu }\right\} \partial _{\mu
}+\left\{ x_{i}^{\mu }((\partial _{\mu }H_{i})-(\partial _{\mu
}R_{i}))+H_{i}H_{i}-R_{i}R_{i}\right\} ).  \label{TT}
\end{eqnarray}%
Because of $\mathbf{H=}H\mathbf{\mathbf{n}}$ and $\mathbf{n=r}^{\xi }\times 
\mathbf{r}^{\zeta }/\sqrt{g}$, i.e., $\mathbf{H}\cdot \mathbf{r}^{\mu
}=0,(\mu =\xi ,\zeta )$, the term in the first parenthesis $\left\{
{}\right\} $ in (\ref{TT}) vanishes. However, if $R_{i}=0$, i.e., the
operator-ordering factors $f_{i}$ are equal to constant, the terms in the
second parenthesis $\left\{ {}\right\} $ in (\ref{TT}) have nonzero
contribution that is $-H^{2}$. To see this fact, we need to use the
Weingarten formula $\partial _{\mu }\mathbf{n}\equiv \mathbf{n}_{\mu
}=-b_{\mu \upsilon }\mathbf{r}^{\upsilon }$ \cite{smirnov} and a relation $%
\mathbf{r}^{\mu }\cdot \partial _{\mu }\mathbf{n=}-\mathbf{r}^{\mu }\cdot 
\mathbf{r}^{\upsilon }b_{\mu \upsilon }=-g^{\mu \upsilon }b_{\mu \upsilon
}=-2H$ \cite{smirnov}. Then $\mathbf{H}$-dependent term in the second
parenthesis $\left\{ {}\right\} $ in (\ref{TT}) is then $x_{i}^{\mu
}(\partial _{\mu }H_{i})+H_{i}H_{i}=$ $\mathbf{r}^{\mu }\cdot (\partial
_{\mu }\mathbf{H})+\mathbf{H\cdot H}=-H^{2}$. So, if $f_{i}=const.$, i.e., $%
\mathbf{R=}0$,\textbf{\ }the result (\ref{wrong}) holds. However, the
presence of the operator-ordering terms $R_{i}$ may cancel out the excess
terms, making the terms in the second parenthesis $\left\{ {}\right\} $ in (%
\ref{TT}) vanish. This requirement leads to the following equation in vector
form, 
\begin{equation}
\mathbf{r}^{\mu }\cdot ((\partial _{\mu }\mathbf{H})-(\partial _{\mu }%
\mathbf{R}))+(H_{i}H_{i}-R_{i}R_{i})=0.  \label{eqset}
\end{equation}%
It is a nonlinear differential equation and trivial case $f_{i}=const.$, $%
\mathbf{R=}0$,\textbf{\ }can never solve it unless $H=0$. A particular
solution for $\mathbf{R}$ is evidently, 
\begin{equation}
\mathbf{R}=\mathbf{H}=H\mathbf{\mathbf{n}.}  \label{RH}
\end{equation}%
When the motion is constraint-free or in a flat plane, i.e., when $H=0$, the
factors $f_{i\text{ \ }}$become trivial for $f_{i}=const$. from Eq. (\ref%
{hdepend}).

\section{Other two operator-orderings in kinetic operator}

In our previous concrete approach \cite{liu1}, we use the following form of
the kinetic operator,%
\begin{equation}
T1=\frac{1}{2m}\overset{3}{\underset{i=1}{\sum }}\frac{1}{f_{i}(x,y,z)}%
p_{i}f_{i}(x,y,z)p_{i}.  \label{ordering2}
\end{equation}%
It is also tempted to use,%
\begin{equation}
T2=\frac{1}{2m}\overset{3}{\underset{i=1}{\sum }}p_{i}f_{i}(x,y,z)p_{i}\frac{%
1}{f_{i}(x,y,z)}.  \label{ordering3}
\end{equation}%
The operator-ordering problem presenting in Eqs. (\ref{ordering2})\ and (\ref%
{ordering3}) differs from the Eq. (\ref{general}) only in the way of
distribution of the operator-ordering factors $f_{i}(x,y,z)$. Next, we prove
that these factors $f_{i}(x,y,z)$ have exactly the same form as it is given
by Eq. (\ref{RH}).

Expanding the right hand side of Eqs. (\ref{ordering2})\ and (\ref{ordering3}%
), we find respectively, 
\begin{eqnarray}
T1 &=&-\frac{\hbar ^{2}}{2m}(x_{i}^{\mu }\partial _{\mu
}+H_{i}+R_{i})(x_{i}^{\upsilon }\partial _{\upsilon }+H_{i})  \notag \\
&=&-\frac{\hbar ^{2}}{2m}(x_{i}^{\mu }\partial _{\mu }x_{i}^{\upsilon
}\partial _{\upsilon }+\left\{ (2H_{i}+R_{i})x_{i}^{\upsilon }\right\}
\partial _{\upsilon }+\left\{ x_{i}^{\mu }(\partial _{\mu
}H_{i})+R_{i}H_{i}+H_{i}H_{i}\right\} ),  \label{ordering21}
\end{eqnarray}%
and,%
\begin{eqnarray}
T2 &=&-\frac{\hbar ^{2}}{2m}(x_{i}^{\mu }\partial _{\mu
}+H_{i})(x_{i}^{\upsilon }\partial _{\upsilon }+H_{i}-R_{i})  \notag \\
&=&-\frac{\hbar ^{2}}{2m}(x_{i}^{\mu }\partial _{\mu }x_{i}^{\upsilon
}\partial _{\upsilon }-\left\{ R_{i}x_{i}^{\mu }\right\} \partial _{\mu
}+\left\{ x_{i}^{\mu }((\partial _{\mu }H_{i})-(\partial _{\mu
}R_{i}))+H_{i}(H_{i}-R_{i})\right\} ).  \label{ordering31}
\end{eqnarray}%
This requirement that the terms in two parenthesis $\left\{ {}\right\} $ in (%
\ref{ordering21}) vanish simultaneously leads to a set of two equations, 
\begin{equation}
\left\{ 
\begin{array}{c}
\mathbf{R}\cdot \mathbf{r}^{\mu }=0,(\mu =\xi ,\zeta ) \\ 
-H^{2}+\mathbf{R}\cdot \mathbf{H}=0%
\end{array}%
\right. .  \label{set1}
\end{equation}%
The same requirement for (\ref{ordering31}) leads to another set of two
equations, 
\begin{equation}
\left\{ 
\begin{array}{c}
\mathbf{R}\cdot \mathbf{r}^{\mu }=0,(\mu =\xi ,\zeta ) \\ 
\mathbf{r}^{\mu }\cdot (\partial _{\mu }\mathbf{H})-(\partial _{\mu }\mathbf{%
R})+\mathbf{H}\cdot (\mathbf{H}-\mathbf{R})=0%
\end{array}%
\right. .  \label{set2}
\end{equation}%
The first equation in either set (\ref{set1}) or (\ref{set2}) $\mathbf{R}%
\cdot \mathbf{r}^{\mu }=0$ states nothing but a fact that the direction of $%
\mathbf{R}$ is along the normal $\mathbf{n}$. The second equation in either
set (\ref{set1}) or (\ref{set2}) determines the magnitude of $\mathbf{R}$,
and the unique solution is $R=H$\textbf{.}

\section{Examples}

In this section, two ideal quantum dots, the spheroidal surface \cite%
{nanophys1} and the toroidal surface \cite{nanophys2}, will be utilized to
illustrate the abstract results developed above.

\subsection{Operators on the spheroidal surface}

The spheroidal surface is with two local coordinates $\theta \in \lbrack
0,2\pi ),\varphi \in \lbrack 0,2\pi )$,%
\begin{equation}
\mathbf{r}=(x,y,z)=(a\sin \theta \cos \varphi ,a\sin \theta \sin \varphi
,b\cos \theta ),
\end{equation}%
where $a$ and $b$ denote two distinct axes. The convariant derivatives $%
\mathbf{r}_{\mu }$ and contravariant derivatives $\mathbf{r}^{\mu }$ can be
easily computed and the results are respectively,%
\begin{equation}
\left( 
\begin{array}{c}
\mathbf{r}_{_{\theta }} \\ 
\mathbf{r}_{_{\varphi }}%
\end{array}%
\right) =\left( 
\begin{array}{lll}
a\cos \theta \cos \varphi , & a\cos \theta \sin \varphi , & -b\sin \theta \\ 
-a\sin \theta \sin \varphi , & a\sin \theta \cos \varphi , & 0%
\end{array}%
\right) ,
\end{equation}%
\begin{equation}
\left( 
\begin{array}{c}
\mathbf{r}^{\theta }\equiv g^{\theta \upsilon }\mathbf{r}_{\upsilon } \\ 
\mathbf{r}^{\varphi }\equiv g^{\varphi \upsilon }\mathbf{r}_{\upsilon }%
\end{array}%
\right) =\frac{1}{a}\left( 
\begin{array}{lll}
G(a,b,\theta )\cos \theta \cos \varphi , & G(a,b,\theta )\cos \theta \sin
\varphi , & -G(a,b,\theta )b/a\sin \theta \\ 
-\csc \theta \sin \varphi , & \csc \theta \cos \varphi , & 0%
\end{array}%
\right) ,
\end{equation}%
where $G(a,b,\theta )=2/\left( 1+\varepsilon ^{2}+\left( 1-\varepsilon
^{2}\right) \cos 2\theta \right) $ with $\varepsilon =b/a$. The normal $%
\mathbf{n}$ and the mean curvature $H$ are given by respectively,%
\begin{equation}
\mathbf{n=}\sqrt{G(a,b,\theta )}(\varepsilon \sin \theta \cos \varphi
,\varepsilon \sin \theta \sin \varphi ,\cos \theta ),
\end{equation}%
\begin{equation}
H(a,b,\theta )=-b/(4a^{2})\left( 3+\varepsilon ^{2}+\left( 1-\varepsilon
^{2}\right) \cos 2\theta \right) G(a,b,\theta )^{3/2}.
\end{equation}%
The Hermitian Cartesian momentum operators $p_{i}$ $(i=1,2,3)$ are,%
\begin{eqnarray}
p_{x} &=&-i\hbar \frac{1}{a}(\cos \theta \cos \varphi G(a,b,\theta )\frac{%
\partial }{\partial \theta }-\csc \theta \sin \varphi \frac{\partial }{%
\partial \varphi }-F(a,b,\theta )\cos \varphi \sin \theta ), \\
p_{y} &=&-i\hbar \frac{1}{a}(\cos \theta \sin \varphi G(a,b,\theta )\frac{%
\partial }{\partial \theta }+\cos \varphi \csc \theta \frac{\partial }{%
\partial \varphi }-F(a,b,\theta )\sin \theta \sin \varphi ), \\
p_{z} &=&i\hbar (\frac{b}{a^{2}}\sin \theta G(a,b,\theta )\frac{\partial }{%
\partial \theta }+\frac{1}{b}F(a,b,\theta )\cos \theta ),
\end{eqnarray}%
where $F(a,b,\theta )=\varepsilon ^{2}\left( 3+\varepsilon ^{2}+\left(
1-\varepsilon ^{2}\right) \cos 2\theta \right) G(a,b,\theta )^{2}/4$. The
factor functions $(f_{x},f_{y},f_{z})$ determined by equation $R_{i}=Hn_{i}$
(\ref{hdepend}) have special solutions:%
\begin{eqnarray}
f_{x} &=&G(a,b,\theta )^{1/4}(\cos \theta )^{\frac{{a}^{2}+{b}^{2}}{2{a}^{2}}%
}{,} \\
f_{y} &=&G(a,b,\theta )^{1/4}(\cos \theta )^{\frac{{a}^{2}+{b}^{2}}{2{a}^{2}}%
}{,} \\
f_{z} &=&G(a,b,\theta )^{1/4}\sin \theta .
\end{eqnarray}

When the spheroid becomes a sphere with $a=b$, we have $\varepsilon =1$, $%
G(a,b,\theta )=1$, $F(a,b,\theta )=1$ and $H(a,b,\theta )=-1$. All results
above readily reduce to those for sphere. \cite{liu1}

\subsection{Operators on the toroidal surface}

The toroidal surface is with two local coordinates $\theta \in \lbrack
0,2\pi ),\varphi \in \lbrack 0,2\pi )$, 
\begin{equation*}
\mathbf{r=}((a+b\sin \theta )\cos \varphi ,(a+b\sin \theta )\sin \varphi
,b\cos \theta ),(a>b)
\end{equation*}%
where $a$ and $b$ denote two distinct radii. The convariant derivatives $%
\mathbf{r}_{\mu }$ and contravariant derivatives $\mathbf{r}^{\mu }$ can be
easily computed and the results are respectively,%
\begin{equation}
\left( 
\begin{array}{c}
\mathbf{r}_{_{\theta }} \\ 
\mathbf{r}_{_{\varphi }}%
\end{array}%
\right) =\left( 
\begin{array}{lll}
b\cos \theta \cos \varphi , & b\cos \theta \sin \varphi , & -b\sin \theta \\ 
-(a+b\sin \theta )\sin \varphi , & (a+b\sin \theta )\cos \varphi , & 0%
\end{array}%
\right) ,
\end{equation}%
\begin{equation}
\left( 
\begin{array}{c}
\mathbf{r}^{\theta }\equiv g^{\theta \upsilon }\mathbf{r}_{\upsilon } \\ 
\mathbf{r}^{\varphi }\equiv g^{\varphi \upsilon }\mathbf{r}_{\upsilon }%
\end{array}%
\right) =\left( 
\begin{array}{lll}
\frac{\cos \theta \cos \varphi }{b}, & \frac{\cos \theta \sin \varphi }{b},
& -\frac{\sin \theta }{b} \\ 
-\frac{\sin \varphi }{a+b\sin \theta }, & \frac{\cos \varphi }{a+b\sin
\theta }, & 0%
\end{array}%
\right) .
\end{equation}%
The normal $\mathbf{n}$ and the mean curvature $H$ are given by respectively{%
,}%
\begin{equation}
\mathbf{n=}\left( 
\begin{array}{l}
\sin \theta \cos \varphi ,\sin \theta \sin \varphi ,\cos \theta%
\end{array}%
\right) ,
\end{equation}%
\begin{equation}
H=-\frac{a+2b\sin \theta }{2b\,{\left( a+b\,\sin \theta \right) }}.
\label{Htorus}
\end{equation}%
With use of the above expression for mean curvature $H$ (\ref{Htorus}), the
Hermitian Cartesian momentum operators $p_{i}$ $(i=1,2,3)$ are given by,%
\begin{eqnarray}
p_{x} &=&-i\hbar \left( \frac{\cos \theta \cos \varphi }{b}\frac{\partial }{%
\partial \theta }-\frac{\sin \varphi }{a+b\,\sin \theta \,}\frac{\partial }{%
\partial \varphi }+H\sin \theta \cos \varphi \right) , \\
p_{y} &=&-i\hbar \left( \frac{\cos \theta \sin \varphi }{b}\frac{\partial }{%
\partial \theta }+\frac{\cos \varphi }{a+b\,\sin \theta \,\,}\frac{\partial 
}{\partial \varphi }+H\sin \theta \sin \varphi \right) , \\
p_{z} &=&i\hbar (\frac{\sin \theta }{b}\frac{\partial }{\partial \theta }%
-H\cos \theta ).
\end{eqnarray}

The factor functions $(f_{x},f_{y},f_{z})$ determined by equation $%
R_{i}=Hn_{i}$ (\ref{hdepend}) have special solutions:%
\begin{eqnarray}
f_{x} &=&\left( a+b\sin \theta \right) ^{\frac{1}{2}\,{\frac{{a}^{2}}{{a}%
^{2}-{b}^{2}}}}\left( 1+\sin \theta \right) ^{\frac{1}{4}\,{\frac{a-2\,b}{a-b%
}}}\left( \sin \theta -1\right) ^{\frac{1}{4}\,{\frac{a+2\,b}{a+b}}}, \\
f_{y} &=&\left( a+b\sin \theta \right) ^{\frac{1}{2}\,{\frac{{a}^{2}}{{a}%
^{2}-{b}^{2}}}}\left( 1+\sin \theta \right) ^{\frac{1}{4}\,{\frac{a-2\,b}{a-b%
}}}\left( \sin \theta -1\right) ^{\frac{1}{4}\,{\frac{a+2\,b}{a+b}}}, \\
f_{z} &=&\sqrt{(a+b\sin \,\theta )\sin \,\theta }.
\end{eqnarray}

In an extreme case $a=0$, the torus becomes a sphere of radius $b$, and all
results above also reduce to those for sphere. \cite{liu1}

\section{Remarks and summary}

In classical mechanics for a particle moving on the curved surface $M$
embedded\ in $R^{3}$, the local curved coordinates $(\xi ,\zeta )$ on $M$
and the Cartesian coordinates $(x,y,z)$ in $R^{3}$ seems to play equal roles
in the description of its classical motion, for the results written in these
two coordinate systems are related to each other by coordinate
transformation. On the other hand, in light of the canonical variable,
neither the Cartesian coordinates nor the Cartesian momentum can be taken as
canonical variables. Any pair of Cartesian variables $(x_{i},p_{i})$ is no
longer canonical conjugate to each other. Even looking for the canonically
conjugate variables for these Cartesian variable $x_{i},p_{i}$ seems not a
physically meaningful task. In contrast, since the variables canonically
conjugate to the local coordinate variables $(\xi ,\zeta )$ naturally exist,
the quantization based on the conjugate variables can be easily preformed
with help of the so-called canonical quantization rules. However, though so
far quantum mechanics uses the local coordinate system only, it contains
nice results associated with the Cartesian coordinates.

The present work shows a compact and abstract result for Hermitian Cartesian
momentum operators describing the particle moving on the curved surface $M$
embedded\ in $R^{3}$, and it is a constant factor $-i\hbar $ times the mean
curvature vector field $H\mathbf{n}$ added to the usual differential $%
\mathbf{r}^{\mu }\partial _{\mu }$. With use of this Cartesian momentum, the
same operator-ordering factors can be distributed in three different ways,
and all lead to the correct quantum kinetic energy. These operator-ordering
factors become dummy in classical limit and reduce to be constant for the
motion is constraint-free or in the flat plane. Thus, the present study
demonstrates that the Cartesian coordinates is also useful in quantum
mechanics, and casts a new insight into the understanding of the classical
correspondence of quantum mechanics \cite{liu6,liu7,liu8}.

\textbf{Acknowledgments}

This Project Supported by Program for New Century Excellent Talents in
University, Ministry of Education, and Student Innovation Training (SIT)
Program, Hunan University, China.

\end{document}